\documentclass[graybox]{svmult}

\usepackage{graphicx}
\usepackage{epsfig}
\usepackage{slashed}

\newcommand{\bel}[1]{\begin{equation}\label{#1}}
\newcommand{\bal}[1]{\begin{eqnarray}\label{#1}}

\newcommand{\be}{\begin{equation}}
\newcommand{\ee}{\end{equation}}
\newcommand{\ba}{\begin{eqnarray}}
\newcommand{\ea}{\end{eqnarray}}

\newcommand{\bes}{\begin{equation*}}
\newcommand{\ees}{\end{equation*}}

\usepackage{latexsym}
\usepackage{amsmath}
\usepackage{amssymb}
\usepackage{eufrak}
\usepackage{euscript}
\usepackage{epsfig,graphics,graphicx}
\usepackage{bm}
\usepackage{color}
\usepackage{feynmf}
 \usepackage{slashed}

\def\be{\begin{eqnarray}}
\def\ee{\end{eqnarray}}

\newcommand{\beqn}{\begin{eqnarray}}
\newcommand{\eeqn}{\end{eqnarray}}

\newcommand{\M}{PLSM${}_q$}

\newcommand{\dd}{{\mathrm d}}

\newcommand{\Tr}{{\mathrm{Tr}\,}}
\newcommand{\Z}{{\mathbb{Z}}}

\usepackage{mathptmx}       
\usepackage{helvet}         
\usepackage{courier}        
\usepackage{type1cm}        
\usepackage{makeidx}         
\usepackage{graphicx}        
\usepackage{multicol}        
\usepackage[bottom]{footmisc}

\makeindex             

\begin{document}

\title*{Thermal chiral and deconfining transitions in the presence of a magnetic background}
\author{Eduardo S. Fraga}
\institute{Eduardo S. Fraga \at Instituto de F\'\i sica, Universidade Federal do Rio de Janeiro, 
Caixa Postal 68528, Rio de Janeiro, RJ 21941-972, Brazil, \email{fraga@if.ufrj.br}}
%
%
\maketitle

\abstract{We review the influence of a magnetic background on the phase diagram of strong 
interactions and how the chiral and deconfining transitions can be affected. First we summarize results 
for both transitions obtained in the framework of the linear sigma model coupled to quarks and to the Polyakov 
loop, and how they compare to other effective model approaches and to lattice QCD. Then we discuss the 
outcome of the magnetic MIT bag model that yields a behavior for the critical deconfining temperature which 
is compatible with recent lattice results and magnetic catalysis. The qualitative success of the magnetic 
MIT bag model hints to $T_{c}$ being a confinement-driven quantity, and leads us to the discussion of 
its behavior as predicted within the large-$N_{c}$ limit of QCD, which is also in line with the most recent 
lattice QCD results provided that quarks behave paramagnetically.}

\section{Introduction}

The thermodynamics of strong interactions under a strong magnetic background has proven to 
be a very rich and subtle subject. Recent developments were initially motivated by the utility of 
magnetic fields in separating charge in space, which would render the possible formation of 
sphaleron-induced CP-odd domains in the plasma created in high-energy heavy ion collisions, 
in the so-called chiral magnetic effect \cite{CME}, measurable. In fact, the magnetic fields created 
in non-central collisions in heavy ion experiments at RHIC-BNL and the LHC-CERN are possibly 
the highest since the epoch of the electroweak phase transition, reaching values such as 
$B \sim 10^{19}~$Gauss ($eB \sim6\,m_{\pi}^2$) for peripheral collisions at RHIC \cite{magnetic-HIC} 
and even much higher at the LHC due to the fluctuations in the distribution of protons inside the 
nuclei \cite{Bzdak:2011yy}.

From the theoretical point of view, the non-trivial role played by magnetic fields in the nature of phase 
transitions has been known for a long time \cite{landau-book}. Modifications in the vacuum of QED and 
QCD have also been investigated within different frameworks, mainly using effective 
models \cite{Klevansky:1989vi,Gusynin:1994xp,Babansky:1997zh,Klimenko:1998su,
Semenoff:1999xv,Goyal:1999ye,Hiller:2008eh,Ayala:2004dx}, 
especially the NJL model \cite{Klevansky:1992qe}, and chiral perturbation 
theory \cite{Shushpanov:1997sf,Agasian:1999sx,Cohen:2007bt}, but also resorting to the 
quark model \cite{Kabat:2002er} and certain limits of QCD \cite{Miransky:2002rp}. 
Interesting phases in dense systems \cite{Ferrer:2005vd,Fukushima:2007fc,Noronha:2007wg,Son:2007ny}, 
as well as effects on 
the dynamical quark mass \cite{Klimenko:2008mg} were also considered. Nevertheless, the 
mapping of the new $T-eB$ phase diagram is still an open problem. There are clear indications that 
sufficiently large magnetic 
fields could significantly modify the behavior of the chiral and the deconfinement phase 
transition lines 
\cite{Agasian:2008tb,Fraga:2008qn,Mizher:2008hf,Mizher:2010zb,Menezes:2008qt,Boomsma:2009yk,Fukushima:2010fe,Johnson:2008vna,Preis:2010cq,Callebaut:2011uc,Avancini:2011zz,Kashiwa:2011js,Chatterjee:2011ry,Andersen:2011ip,Andersen:2012dz,Skokov:2011ib,Fraga:2012fs,Fukushima:2012xw}, or even  
transform the vacuum into a superconducting medium via $\rho$-meson 
condensation \cite{Chernodub:2010qx}. Although most of the analyses so far relied on effective 
models, lattice QCD has definitely entered the field and has been producing its first results for the phase 
diagram \cite{lattice-maxim,lattice-delia,Braguta:2011hq,Bali:2011qj,Bali:2012zg}.

\begin{figure}[!thb]
\vskip -4mm
\begin{center}
\includegraphics[width=85mm,clip=true]{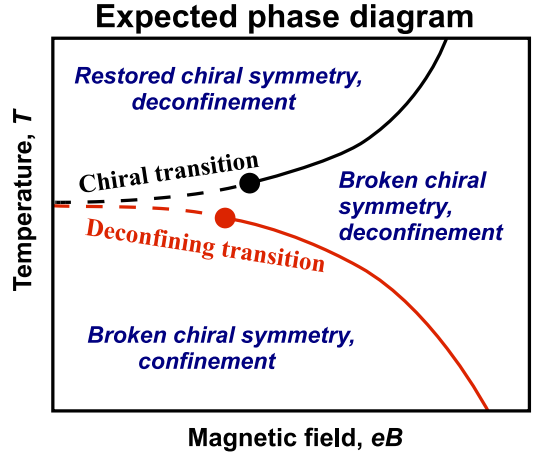}
\end{center}
\vskip -47mm
\caption{Originally expected magnetic field--temperature phase diagram of strong interactions.
The thick lines indicate first-order transitions, the filled circles are the (second-order) endpoints
of these lines, and the thin dashed lines stand for the corresponding crossovers. A new phase with 
broken chiral symmetry and deconfinement appears at high magnetic fields. Extracted from 
Ref. \cite{Mizher:2010zb}.}
\label{fig:expected}
\end{figure}

From the first results obtained within effective models for the deconfining \cite{Agasian:2008tb} and 
chiral \cite{Fraga:2008qn} transition lines, one would expect the phase diagram structure illustrated 
in Fig. \ref{fig:expected}, as discussed in Ref. \cite{Mizher:2010zb}. Indeed, after the prediction of a 
splitting between the chiral and deconfining transition lines, with the appearance of a new phase, 
in Ref. \cite{Mizher:2010zb}, several model descriptions produced the same 
effect \cite{Fukushima:2010fe,Johnson:2008vna,Preis:2010cq,Callebaut:2011uc,Skokov:2011ib}. 
However, until 2011, all model studies 
have yielded either a monotonically increasing or an essentially flat functional form for the 
deconfinement critical line as $B$ increases to very large values\footnote{Contrastingly, a significant 
decrease in the critical temperature as a function of $B$, vanishing at $eB_{c}\sim 25m_{\pi}^2$, 
was found in Ref. \cite{Agasian:2008tb}, featuring the disappearance of the 
confined phase at large magnetic fields. This phenomenon that was not reproduced by any other effective 
model nor observed on the lattice (even for much larger fields).}. Pioneering lattice 
simulations  \cite{lattice-delia} also found an essentially flat behavior for both critical lines, that seemed 
to increase together at a very low rate. Nevertheless, since the pion mass used in these simulations 
was still very high, this could be an indication that one would probably need huge magnetic fields in 
the simulations in order to be able to compare to effective model predictions.

This was the scenario, rather coherent in terms of expectations for the behavior of the critical lines 
for the chiral and deconfining transitions in the presence of a magnetic background, until lattice 
simulations of magnetic QCD with physical masses and fine grids were performed \cite{Bali:2011qj} 
and showed that both critical temperatures actually go down for increasing $B$, saturating 
for very large fields, very differently from what has been predicted by all previous effective model 
calculations and found in previous lattice simulations. 

Soon after the appearance of the new lattice results, the behavior of the critical temperature for 
deconfinement in the presence of a very large magnetic field was addressed within the 
MIT bag model \cite{Fraga:2012fs}, a very economic model in 
terms of parameters to be fixed (essentially one) and other ingredients usually hard to control in more 
sophisticated effective theories. The model is, of course, crude in numerical precision and misses the 
correct nature of the (crossover) transition. Nevertheless, it provides a simple setup for the discussion 
of some subtleties of vacuum and thermal contributions in each phase. It was shown in 
Ref. \cite{Fraga:2012fs} that the influence of the magnetic field on the thermodynamics of both extreme 
energy domains is captured, so that the model furnishes a reasonable qualitative description of the 
behavior of the critical temperature in the presence of $B$, decreasing and saturating. 

The fact that chiral models, even when coupled with the (static) Polyakov loop sector, seem to fail 
in the description of the behavior of $T_{c} \times eB$, whereas the (assumedly simple) MIT bag 
approach finds a good qualitative agreement, suggests that the critical temperature in QCD is a 
confinement-driven observable. This was also hinted by a previous successful description of the behavior 
of the critical temperature as a function of the pion mass and isospin chemical potential, as compared to 
lattice data, where chiral models also failed even qualitatively \cite{Fraga:2008be,leticia-tese}.
If confinement dynamics plays a central role in guiding the functional behavior of $T_{c}$, a the large-$N_{c}$ 
limit of QCD should provide an adequate and powerful framework to study associated magnetic 
thermodynamics. In fact, it was shown in Ref. \cite{mag-largeNc} on very general grounds that the fact that 
the deconfining temperature decreases and tends to saturate for large $B$, although this last point 
cannot be proven in a model-independent way, depends solely on quarks behaving paramagnetically.

In the sequel we summarize results for the chiral and deconfining transitions obtained in the 
framework of the linear sigma model coupled to quarks and to the Polyakov loop, especially the 
prediction of a splitting of the two critical lines, and how they compare to other effective model 
approaches as well as to lattice QCD. Then we discuss the outcome of the magnetic MIT bag model 
that yields a behavior for the critical deconfining temperature compatible with the most recent lattice 
simulations and magnetic catalysis. We continue with a discussion of very recent results, starting 
with the rather general findings within the large-$N_{c}$ limit of magnetic QCD. Finally, we present 
our conclusions. 

\section{Modified dispersion relations and integral measures}

In the presence of a classical, constant and uniform (Abelian) magnetic field, dispersion relations 
and momentum integrals will be modified. In order to compute vacuum and thermal determinants 
and Feynman diagrams, it is necessary to express these quantities in a convenient fashion. Lorentz 
invariance is broken by the preferred direction established by the external field, and Landau orbits 
redefine the new counting of quantum states \cite{landau-book}. 

For definiteness, let us take the direction of the magnetic field as the $z$-direction, $\vec{B}=B {\bf \hat z}$. 
One can compute, for instance, the modified effective potential or the modified  pressure to lowest order 
by redefining the dispersion relations of charged scalar and spinor fields in the presence of $\vec{B}$, 
using the minimal coupling shift in the gradient and the field equations of 
motion\footnote{Higher-order (loop) corrections need the full propagator, not only its poles.}. 
For this purpose, 
it is convenient to choose the gauge such that $A^{\mu}=(A^{0},\vec{A})=(0,-By,0,0)$.

For scalar fields with electric charge $q$, such as pions, one has
\begin{eqnarray}
&&(\partial^2 +m^2)\phi=0 \, ,\\
&&\partial_\mu \to \partial_\mu+iq A_\mu \, .
\end{eqnarray}
After decomposing $\phi$ into Fourier modes, except for the dependence in the 
coordinate $y$, one obtains
\begin{eqnarray}
\varphi''(y)&+&2m\left[ \left( \frac{p_0^2-p_z^2-m^2}{2m} \right) 
-\frac{q^2B^2}{2m}\left( y+\frac{p_x}{qB} \right)^2\right]\varphi(y)=0 \, ,
\end{eqnarray}
which has the form of a Schr\"odinger equation for a harmonic oscillator. Its eigenmodes 
correspond to the well-known Landau levels
\begin{eqnarray}
\varepsilon_n &\equiv& \left( \frac{p_{0n}^2-p_z^2-m^2}{2m} \right)=
\left( \ell+\frac{1}{2} \right)\omega_B \, ,
\end{eqnarray}
where $\omega_B = |q|B/m$ and $\ell$ is a positive ($\ell \geq 0$) integer, and provide the new dispersion relation:
\begin{eqnarray}
p_{0n}^2=p_z^2+m^2+(2\ell+1)|q|B \, .
\end{eqnarray}

One can proceed in an analogous way for fermions with charge $q$. From the free Dirac 
equation $(i \gamma^\mu\partial_\mu -m)\psi=0$, and the shift in $\partial_{\mu}$, 
one arrives at the following Schr\"odinger equation
\begin{eqnarray}
u_{s}''(y)&+&2m \left[ \left( \frac{p_0^2-p_z^2-m^2+|q|Bs}{2m} \right) 
-\frac{q^2B^2}{2m}\left( y+\frac{p_x}{qB} \right)^2\right]u_{s}(y)=0 \, ,
\end{eqnarray}
which yields the new dispersion relation for quarks:
\begin{eqnarray}
p_{0n}^2=p_z^2+m^2+(2\ell+1-s)|q|B \, ,
\end{eqnarray}
where $s=\pm 1$ is the spin projection in the ${\bf \hat z}$ direction.

It is also straightforward to show that integrals over four momenta and thermal 
sum-integrals acquire the following forms, respectively \cite{Fraga:2008qn,Chakrabarty:1996te}:
\begin{eqnarray}
&&\int \frac{d^4k}{(2\pi)^4} \mapsto \frac{|q|B}{2\pi}\sum_{\ell=0}^\infty 
\int \frac{dk_0}{2\pi}\frac{dk_z}{2\pi} \, ,\\
T\sum_{n} &&\int \frac{d^3k}{(2\pi)^3} \mapsto \frac{|q|BT}{2\pi}\sum_n \sum_{\ell=0}^\infty 
\int \frac{dk_z}{2\pi} \, ,
\end{eqnarray}
where $\ell$ represents the different Landau levels and $n$ stands for the Matsubara 
frequency indices \cite{kapusta-gale}.

\section{\M\ effective model and the splitting of the chiral and deconfining transition lines}

Let us consider the two-flavor linear sigma model coupled to quarks and to the Polyakov loop, 
the \M\ effective model, in the presence of an external magnetic field \cite{Mizher:2010zb}. 

The confining properties of QCD are encoded in the  complex-valued Polyakov loop variable
$L$. As a matter of fact, the Polyakov loop sector only provides a description of the 
behavior of the approximate order parameter for the $Z(3)$ symmetry, which is explicitly broken 
by the presence of quarks. It is convenient for modeling the deconfining transition and has a good 
agreement with lattice results for most thermodynamic quantities such as the pressure and energy density, 
especially for the pure glue theory, but it does not provide a dynamical description of 
confinement\footnote{This will be a key feature in the discussion of recent results for the 
critical temperature, since $T_{c}$ seems to be a confinement-driven observable for both QCD 
transitions.}. 

The expectation value of the Polyakov loop $L$ is an {\it exact} order parameter
for color confinement in the limit  of infinitely massive quarks:
\beqn
\mbox{Confinement}:\;
\left\{
\begin{array}{llll}
\langle L \rangle  & = & 0 \; , \; & \mbox{low $T$}   \\
\langle L \rangle  & \neq & 0 \; , \; & \mbox{high $T$}
\end{array}
\right.\,,
\;\;
L(x) = \frac{1}{3} \Tr {\cal P} \exp \Bigl[i \int\limits_0^{1/T}
\dd \tau A_4(\vec x, \tau) \Bigr]\,,
\label{eq:L}
\label{eq:L:phases}
\eeqn
where $A_4 = i A_0$ is the matrix-valued temporal component of the Euclidean
gauge field $A_\mu$ and the symbol ${\cal P}$ denotes path ordering. The
integration takes place over compactified imaginary time $\tau$, with periodic boundary conditions.

The chiral features of the model are encoded in the dynamics of the $O(4)$
chiral field, which is an exact order parameter in the chiral limit, in which
quarks and pions are massless degrees of freedom:
\beqn
\mbox{Chiral symmetry}:\;
\left\{
\begin{array}{llll}
\langle \sigma \rangle  & \neq & 0 \; , \; & \mbox{low $T$}   \\
\langle \sigma \rangle  & = & 0 \; ,\; & \mbox{high $T$}
\end{array}
\right.\,,
\label{eq:sigma:phases}
\quad
\begin{array}{lll}
\phi & = & (\sigma,\vec{\pi})\,,\\
\vec{\pi} & = & (\pi^{+},\pi^{0},\pi^{-})\,.
\end{array}
\label{eq:phi}
\eeqn
Here $\vec{\pi}$ is the isotriplet of the pseudoscalar pion fields and
$\sigma$ is the chiral scalar field which plays the role of an approximate
order parameter of the chiral transition in QCD, since chiral symmetry is explicitly broken 
by the nonzero quark masses.

Within this effective model, the quark field $\psi$ connects the Polyakov loop $L$ and the chiral field
$\phi$, making a bridge between confining and chiral properties. 
Quarks are also coupled to the external magnetic field
since the $u$ and $d$ quarks are electrically charged. Thus, it is clear
that the external magnetic field will affect the chiral dynamics as well as 
the confining properties of the model, as much as the latter can be captured by the Polyakov loop sector.

This represents a natural generalization of the linear sigma model coupled to 
quarks  \cite{GellMann:1960np}, an effective theory that has been widely used 
to describe different aspects of the chiral transition, such as thermodynamic 
properties \cite{quarks-chiral,ove,Scavenius:1999zc,Caldas:2000ic,Scavenius:2000qd,
Scavenius:2001bb,paech,Mocsy:2004ab,Aguiar:2003pp,Schaefer:2006ds,Taketani:2006zg}
and the nonequilibrium phase conversion process \cite{Fraga:2004hp}. This generalization 
differs from previous ones \cite{polyakov,explosive} by the inclusion of a bridge via the covariant 
derivative and, of course, because of the modifications brought about by the magnetic field.

The Lagrangian of \M\  describes the constituent quarks $\psi$,
which interact with the meson fields $\sigma$, $\pi^\pm = (\pi^1 \pm i \pi^2)/\sqrt{2}$ and $\pi^0 = \pi^3$,
the Abelian gauge field $a_\mu = (a^0,\vec a) = (0, - B y,0,0)$, 
and the $SU(3)$ gauge field $A_\mu$ via the covariant derivative
$D^{(q)}_{\mu} = (\partial _{\mu} - i Q\, a_\mu - i A_\mu)$ with the charge matrix 
$Q = {\mathrm{diag}} (+ 2e/3, -e/3)$. Its explicit form is given by
\beqn
{\cal L} &=&  \overline{\psi} \left[i \gamma^{\mu}D^{(q)}_{\mu} - g(\sigma +i\gamma _{5}
 \vec{\tau} \cdot \vec{\pi} )\right]\psi + \frac{1}{2}\bigl[(\partial _{\mu}\sigma)^2 + (\partial _{\mu} \pi^0)^2\bigr] 
\nonumber \\
&+& |D^{(\pi)}_\mu|^2 - V_\phi(\sigma ,\vec{\pi}) - V_L(L,T)\,,
\label{eq:L:full}
\eeqn
where $D_\mu^{(\pi)} = \partial_\mu + i e a_\mu$ is the covariant derivative acting on colorless pions. 

The chiral potential has the form
\beqn
V_\phi(\sigma ,\vec{\pi}) = \frac{\lambda}{4}(\sigma^{2}+\vec{\pi}^{2} - {\it v}^2)^2-h\sigma\,,  
\eeqn
where $h = f_{\pi} m_{\pi}^2$, $v^2 = f^2_\pi-{m^{2}_{\pi}}/{\lambda}$, $\lambda = 20$, 
$f_\pi \approx 93\,\mbox{MeV}$ and $m_\pi \approx 138\,\mbox{MeV}$.
The constituent quark mass is given by
$m_q \equiv m_q(\langle\sigma\rangle)= g \langle\sigma\rangle$,
and, choosing $g=3.3$ at $T=0$, one obtains for the constituent quarks in the
vacuum $m_q \approx 310~$MeV. At low temperatures quarks are not
excited, and the model reproduces results from the usual linear
$\sigma$-model without quarks.

The Polyakov potential adopted is given by \cite{ref:fukushima,ref:ratti07,ref:ratti08}
\beqn
\frac{V_L(L,T)}{T^4} &=&-\frac{L^*L}{2}\sum_{l=0}^2 a_l \left(\frac{T_0}{T}\right)^l 
\nonumber \\
&+& b_3\left(\frac{T_0}{T} \right)^3\,
\log\left[1-6\,L^*L+4\left({L^*}^3+L^3\right) - 3\left(L^*L\right)^2\right]\,,
\label{eq:VL}
\eeqn
where $T_0 \equiv T_{SU(3)} = 270\, \mbox{MeV}$ is the critical temperature 
in the pure gauge case and $a_0 = 16\,\pi^2/45 \approx 3.51$, $a_1 = -2.47$, $a_2 = 15.2$, and $b_3 = -1.75$.
Below we follow a mean-field analysis in which the
mesonic sector is treated classically whereas quarks represent fast degrees of freedom.

\begin{figure}[!thb]
\begin{center}
\begin{tabular}{cc}
\includegraphics[width=53mm,clip=true]{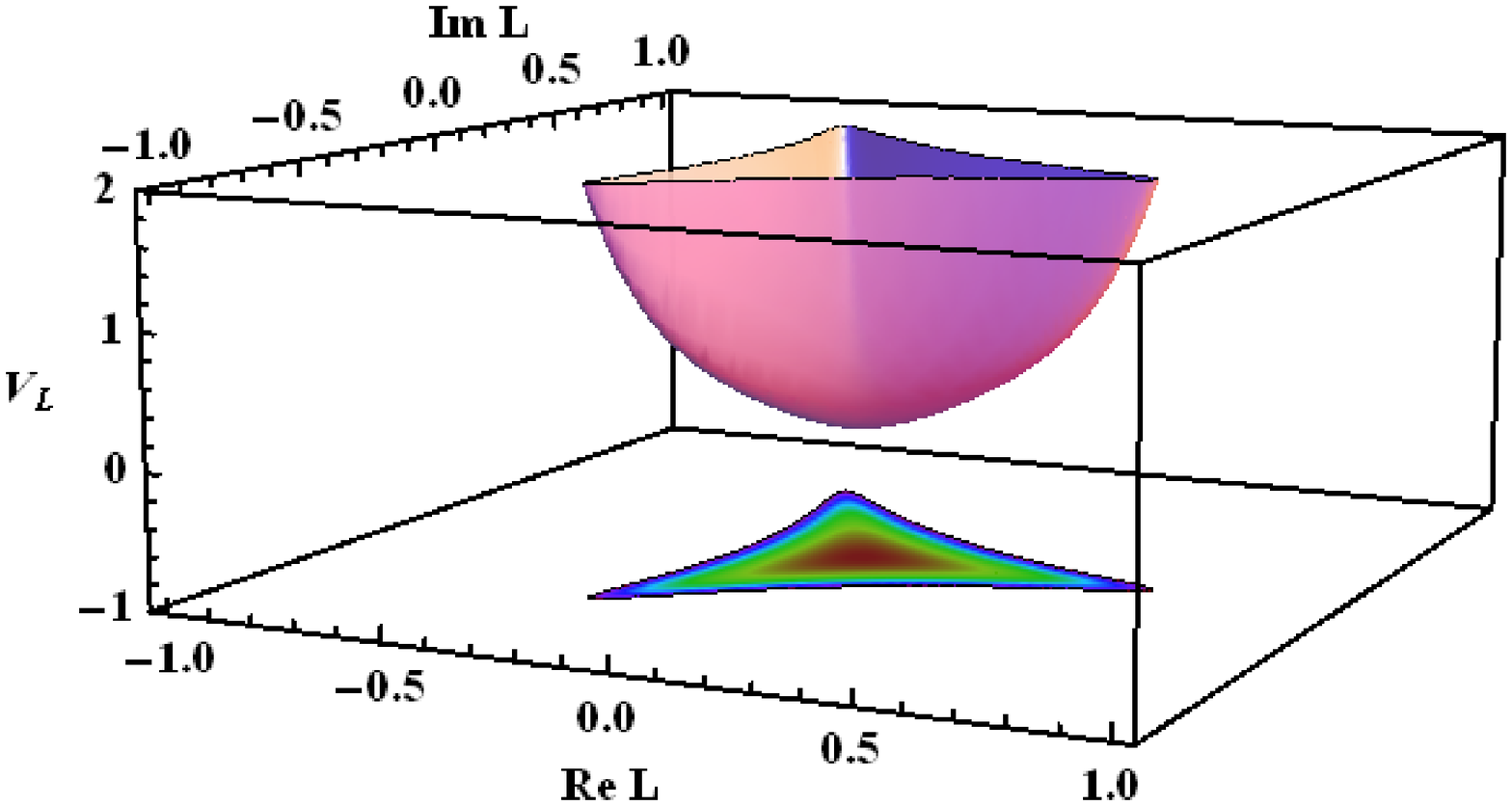} &
\includegraphics[width=55mm,clip=true]{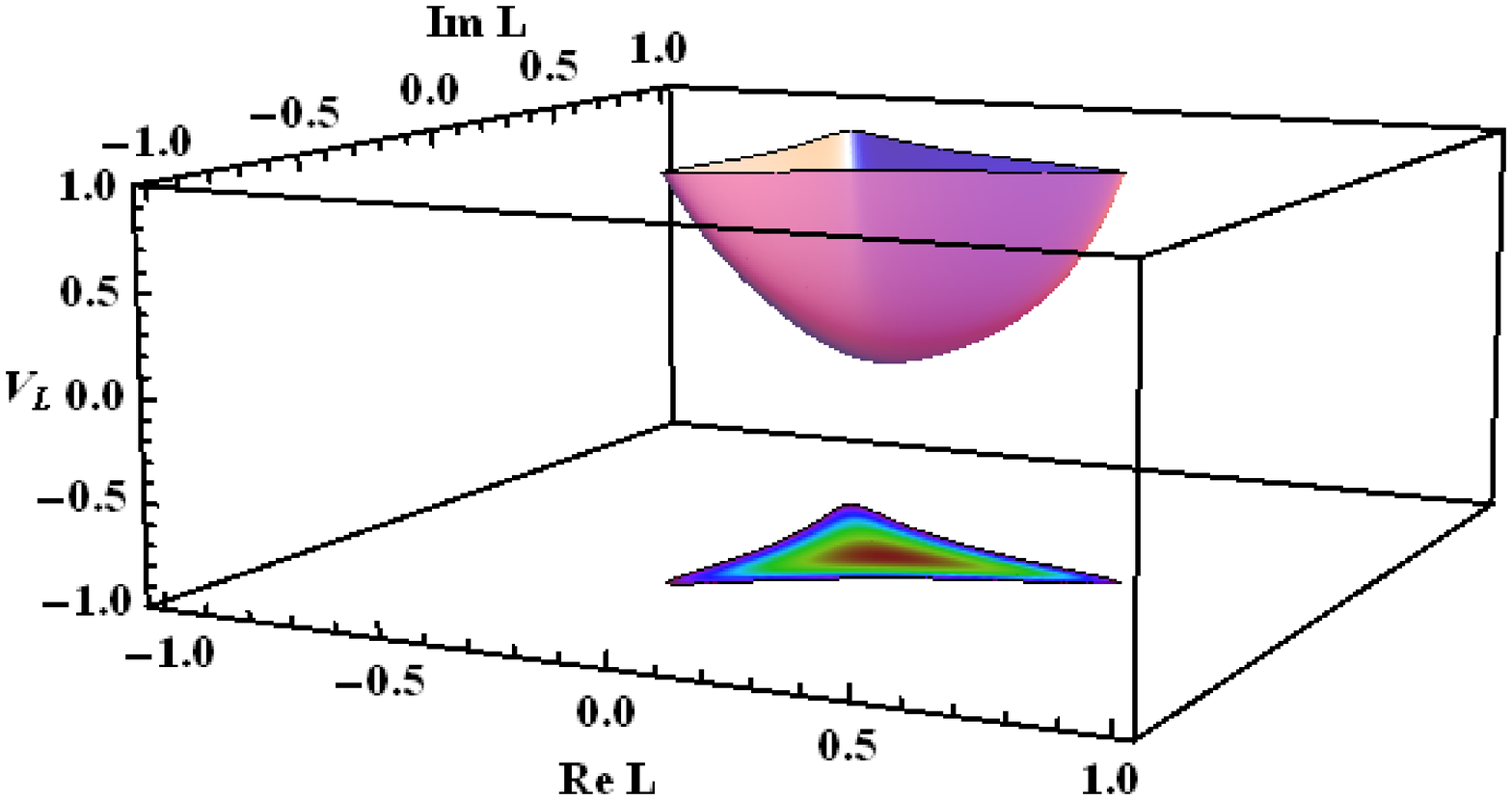} \\[-43mm]
\hskip 15mm {} & 
\hskip 35mm {} \\[43mm]
\includegraphics[width=53mm,clip=true]{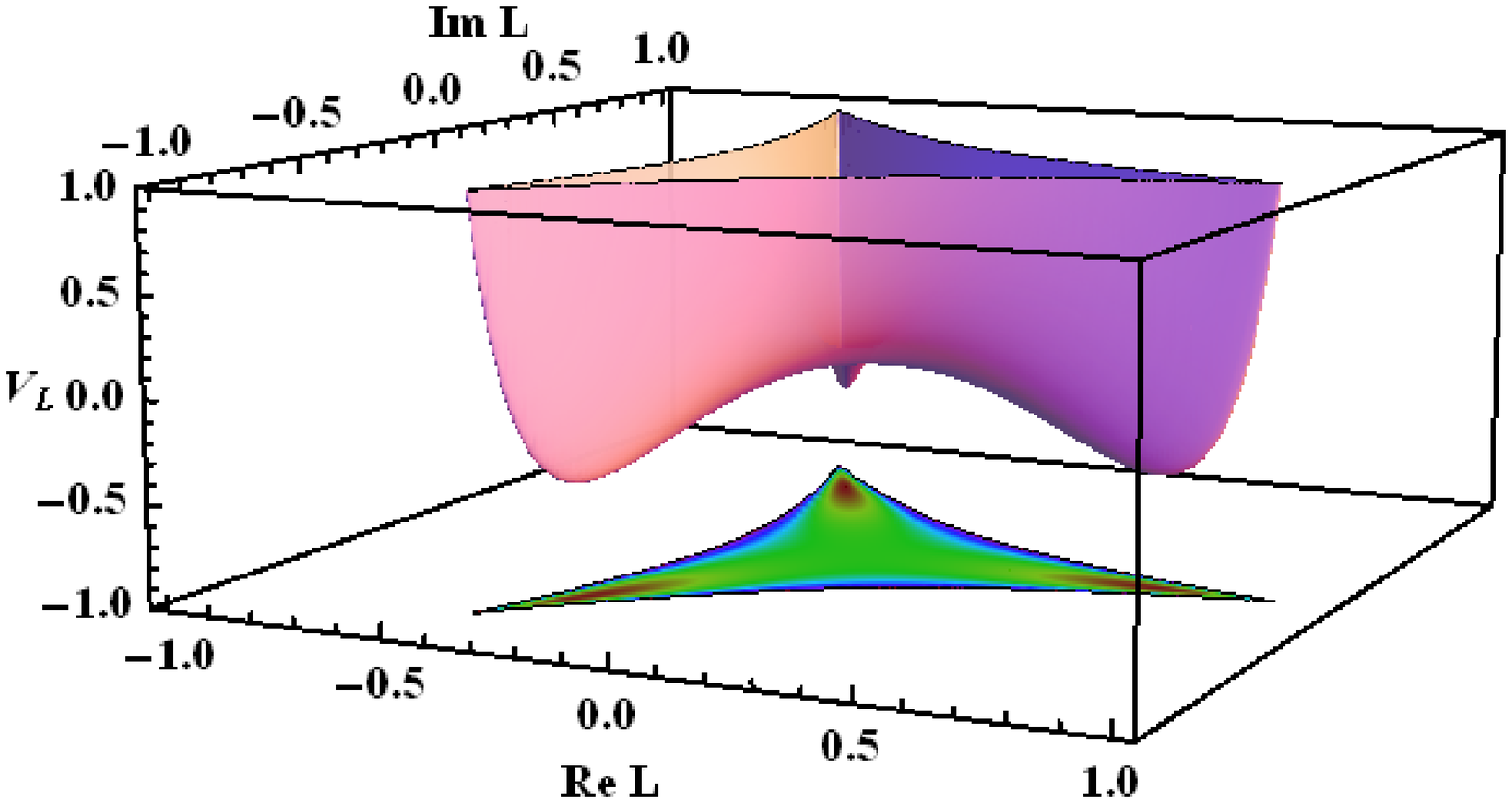} &
\includegraphics[width=55mm,clip=true]{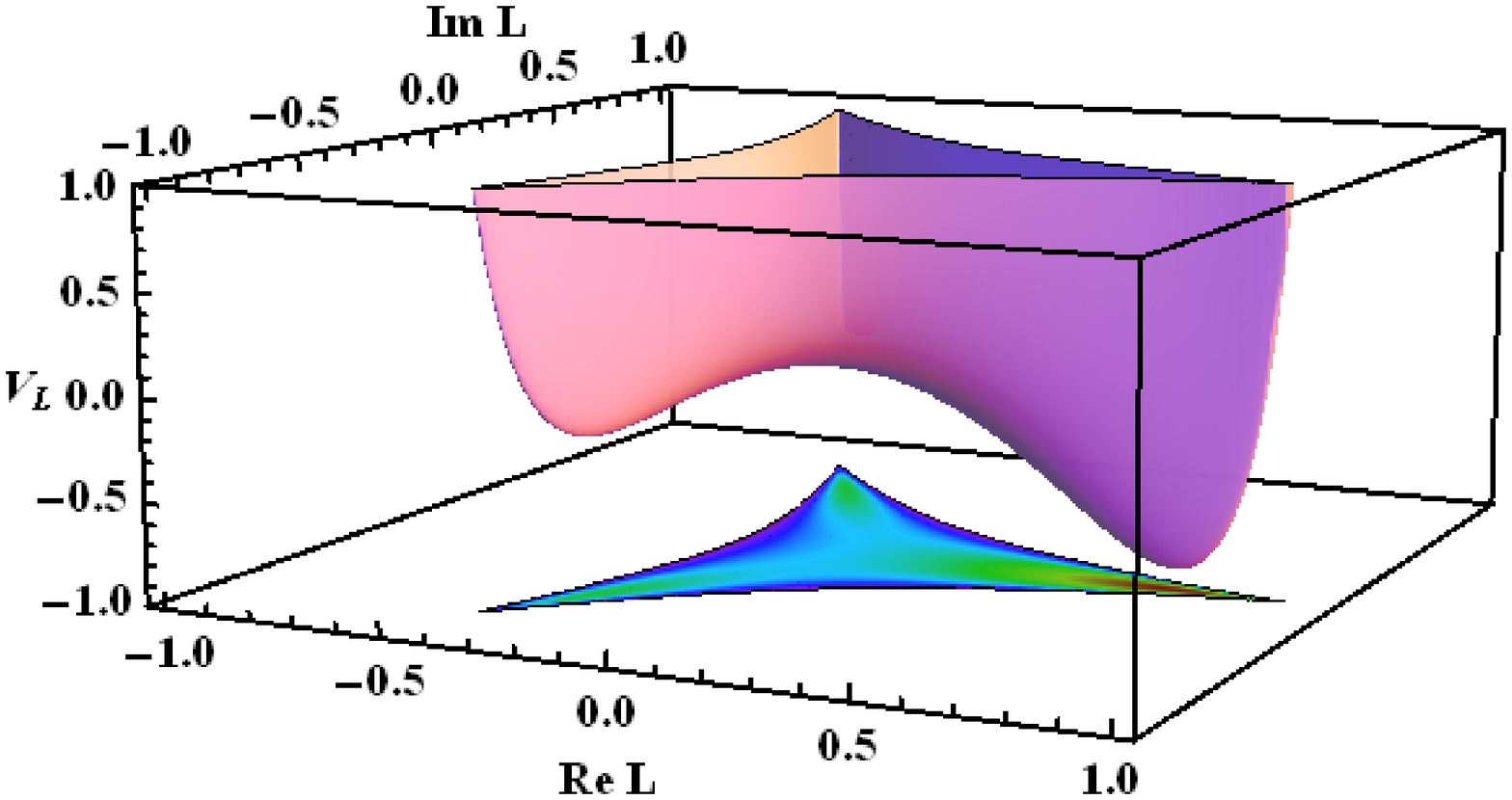} \\[-43mm]
\hskip 35mm {} &
\hskip 35mm {}\\[35mm]
\end{tabular}
\end{center}
\caption{Effects of temperature and magnetic field on quark confinenemt:
The Polyakov loop potential at $T=0.8 \, T_0$ (top) and $T=1.2 \, T_0$ (bottom) and at zero magnetic field (left) and at $e B = 9 T^2$ (right). Extracted from Ref. \cite{Mizher:2010zb}.}
\label{fig:potential}
\end{figure}

The one-loop corrections to the free energy $\Omega$ coming from quarks can be written as:
\beqn
e^{i V_{3d} \, \Omega_q/T}  =
\left[\frac{\det(i \gammaö^{\mu} D_{\mu}^{(q)} - m_q )}{\det (i \gamma^{\mu}\partial_{\mu} - m_q)} \right] 
 \cdot
\left[\frac{\det_T(i \gammaö^{\mu} D_{\mu}^{(q)} - m_q)}{\det (i \gammaö^{\mu} D_{\mu}^{(q)} - m_q )}\right]\,,
\label{eq:Omega:quark2}
\eeqn
so that the expectation values of the condensates can be obtained by minimizing the free energy
\beqn
\Omega(\sigma,L;T,B) & = & V_\phi(\sigma,\vec\pi) + V_{L}(L,T)
+ \Omega_q(\sigma,L,T) \,, 
\label{eq:Omega}
\eeqn
at fixed values of temperature and magnetic field.
The interaction piece $\Omega_q(\sigma,L,T)$ can be split into a vacuum 
(temperature-independent but still magnetic-field dependent) contribution 
and a thermal correction. The vacuum term has the form
\beqn \nonumber
\Omega^{\mathrm{vac}}_{q}(B) = 
-\frac{N_{c}}{\pi}
\sum_{f=u,d}|q_{f}|B \left[ \left(
\sum_{n=\ell}^\infty 
I_{B}^{(1)}(M_{\ell f}^{2})\right)
\right. 
 - \left.
\frac{I_{B}^{(1)}(m_{f})}{2}\right] \,
\eeqn
minus the 
standard
vacuum correction in the absence of the magnetic field,
\beqn
\Omega_{q}^{(0)}&=& 2N_{c}\sum_{f=u,d}I_{B}^{(3)}(m_{f}^{2}) \, ,
\eeqn
where we have defined the integral
\beqn
I_{B}^{(d)}(M^{2})&=& \int \frac{d^{d}p}{(2\pi)^{d}} \sqrt{p^{2}+M^{2}}
\eeqn
and $M^2_{\ell f} = m^2_f + 2 \ell |q_f| B$. The thermal (paramagnetic) contribution is given by 
\cite{Mizher:2010zb}
\beqn
\Omega^{\mathrm{para}}_q & = & \frac{|q_{f}| B T}{\pi^2} 
\sum_{s = \pm \frac{1}{2}} \sum_{\ell=0}^\infty \sum_{k=1}^\infty
\frac{(-1)^k}{k} \mathrm{Re}\, \left[\mathrm{Tr} \Phi^k \right] 
\mu_{s\ell}(\sigma)\,  K_1\left[\frac{k}{T} \mu_{s\ell}(\sigma)\right] \, ,
\label{eq:para}
\eeqn
where $\mu_{s\ell}$ is the energy of the $\ell^{th}$ Landau level at zero longitudinal momentum, 
$\mu_{s\ell}(\sigma)  = {\bigl[g^2\sigma^2 + (2\ell + 1 - 2 s) |q| B\bigr]}^{1/2}$, 
and the untraced Polyakov loop is such that
$\mathrm{Re}\, \left[\mathrm{Tr} \Phi^k \right] = \sum_{i=1}^3 \cos(k \varphi_i)$,  
the integer $k$ corresponding to the winding number of the Polyakov loops \cite{Mizher:2010zb}.

\begin{figure}[!thb]
\vskip 3mm
\begin{center}
\includegraphics[width=75mm,clip=true]{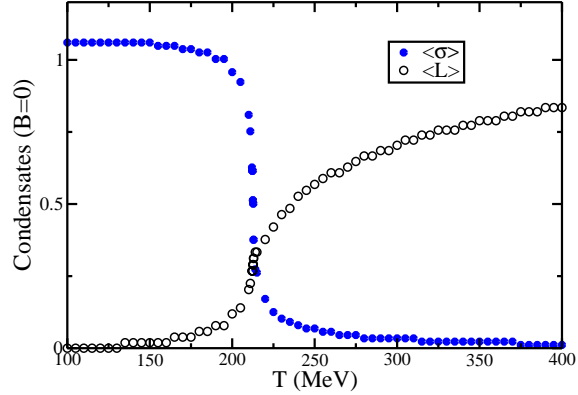}
\end{center}
\vskip -3mm
\caption{
Expectation values
of the order parameters for the chiral and deconfinement transitions as functions of the temperature. 
The filled circles represent the
$\sigma$--condensate, and the empty circles stand for the expectation value of the Polyakov loop. 
In this plot the condensates are dimensionless. Extracted from Ref. \cite{Mizher:2010zb}.}
\label{fig:B0}
\end{figure}

For finite temperature and $B=0$ this model produces a crossover for both transitions. 
Fig. \ref{fig:B0} displays the condensates as functions of the temperature, and the critical temperature 
is defined by the change curvature in the curves. This occurs simultaneously for the chiral and 
deconfinement transitions within this model.

\begin{figure}[!thb]
\begin{center}
\includegraphics[width=85mm,clip=true]{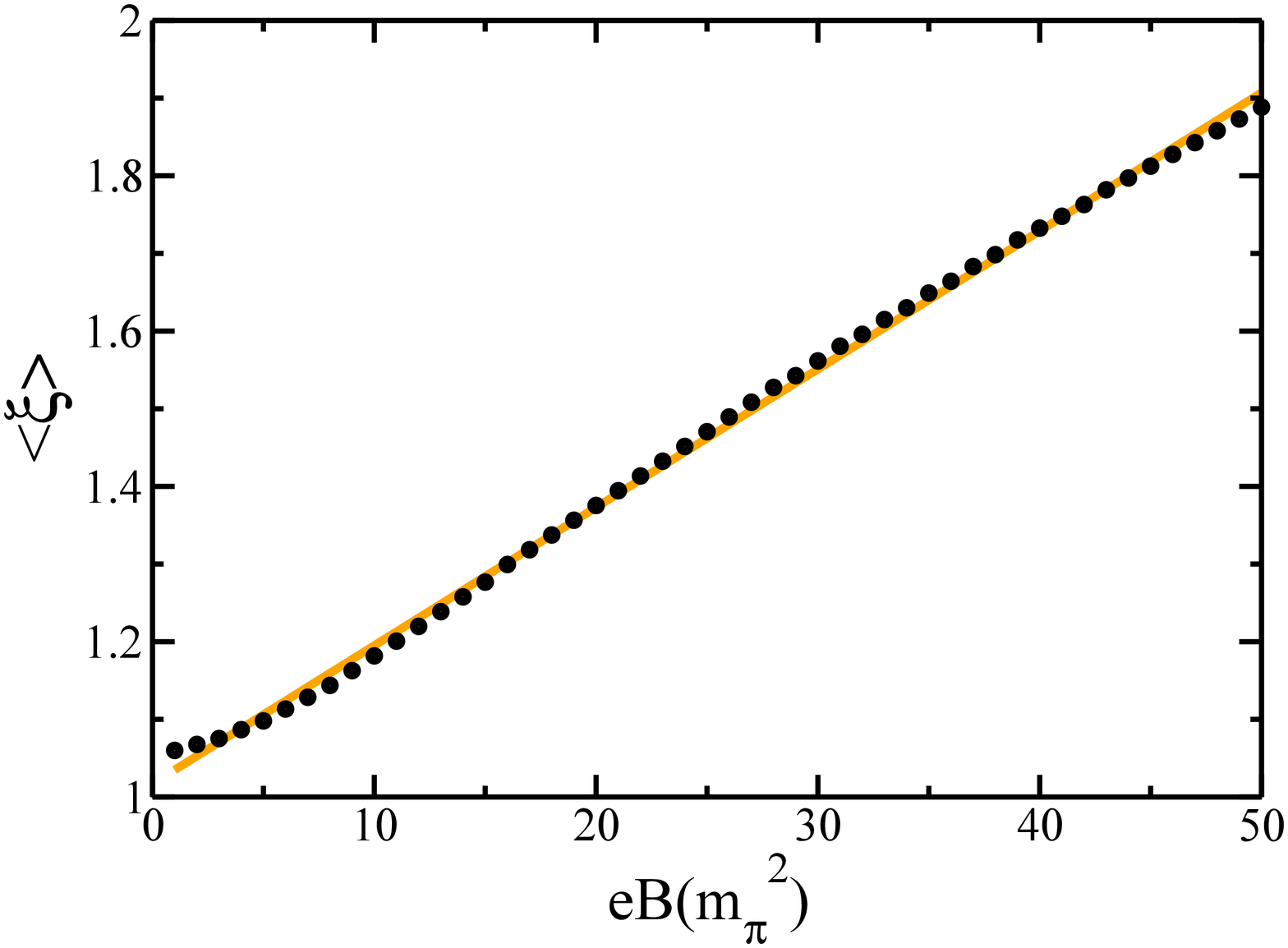}
\includegraphics[width=75mm,clip=true]{zerotemp_severalB.eps}
\end{center}
\caption{Upper: the expectation value of the (dimensionless, $\sigma = \xi v$) 
condensate as a function of the magnetic field. 
Black dots are obtained from the PLSMq and the orange line is the linear fit. 
Lower: effective potential for the condensate 
at zero temperature for several values of the magnetic field $B$. Extracted from Ref. \cite{Mizher:2010zb}.}
\label{fig:ximin_B}
\end{figure}

At zero temperature, the Polyakov loop variable does not play a role. The presence of a magnetic field 
enhances the chiral symmetry breaking, increasing the value of the chiral condensate, in line with the 
phenomenon of magnetic catalysis \cite{Gusynin:1994xp,Klimenko:1998su,Semenoff:1999xv,Miransky:2002rp,Shovkovy:2012zn}. 
This is shown in Fig. \ref{fig:ximin_B}. It also deepens the minimum of the potential as $B$ is increased, 
as illustrated in the same figure for several values of the magnetic field 

The dependence of the chiral condensate on the magnetic field is approximately linear, as shown 
in Fig. \ref{fig:ximin_B}. This is in line with results from chiral perturbation theory \cite{Shushpanov:1997sf,Agasian:1999sx,Cohen:2007bt}. Recent lattice results \cite{Bali:2012zg} 
seem to deviate from a linear behavior for large $B$, growing faster, in better 
qualitative agreement with results from PNJL \cite{Fukushima:2010fe}. However, for larger values of  $B$, 
all model calculations seem to deviate from the lattice data, whereas for very small $B$ they all 
quantitatively agree \cite{Bali:2012zg}.

\begin{figure}[!thb]
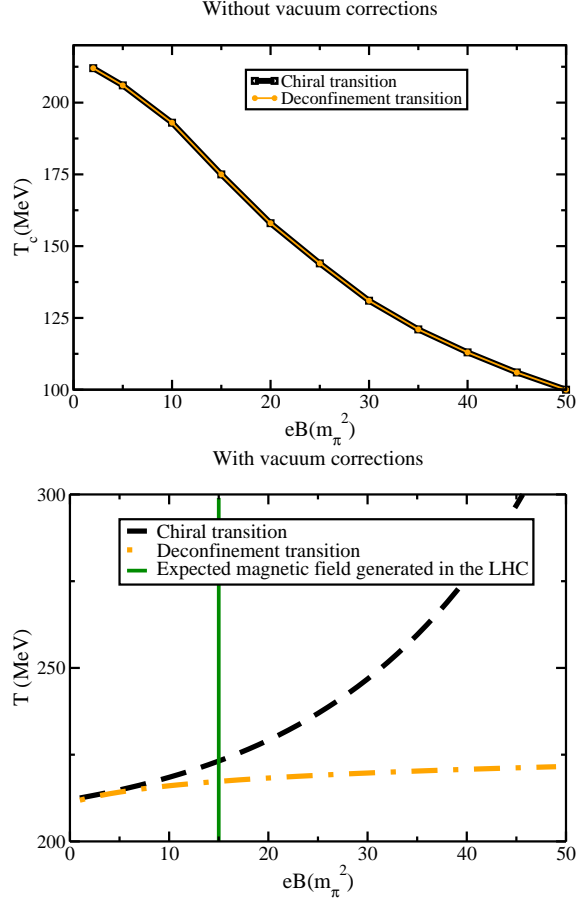

\begin{center}
\includegraphics[width=75mm,clip=true]{nonvacuum_phasedia.eps} 
\includegraphics[width=75mm,clip=true]{vac_phase_diagram.eps}
\end{center}
\caption{Phase diagram in the $B$-$T$ plane. 
Upper: without vacuum corrections: the critical temperatures of the deconfinement (the dash-dotted line) 
and chiral (the dashed line) transition coincide all the way, and decrease with $B$. 
Lower: with vacuum corrections:  the critical temperatures of the deconfinement (the dash-dotted line) 
and chiral (the dashed line) transition coincide at $B=0$ and split
at higher values of the magnetic field. A deconfined phase with broken chiral symmetry appears.
The vertical line is the magnitude of the magnetic field that expected to be realized at LHC heavy-ion collisions~\cite{magnetic-HIC}. Extracted from Ref. \cite{Mizher:2010zb}.}
\label{fig:phase:diagram}		
\end{figure}

Turning on the temperature, one can investigate the effects of the magnetic field on the thermodynamics 
and phase structure of strong interactions as captured by this model description. 
In the confining sector, the strong magnetic field affects the potential for the expectation value 
of the Polyakov loop via the intermediation of the quarks in three ways \cite{Mizher:2010zb}:
(i) the presence of the magnetic field intensifies the breaking of the global $\Z_3$ symmetry and 
makes the Polyakov loop real-valued, as shown in Fig. \ref{fig:potential}; 
(ii) the thermal contribution from quarks tends to destroy the confinement phase by increasing the 
expectation value of the Polyakov loop; 
(iii) on the contrary, the vacuum quark contribution tends to restore the confining phase by 
lowering the expectation value of the Polyakov loop. 

In fact, the vacuum correction from quarks has a crucial impact on the phase structure. If one disregards 
the vacuum contribution from the quarks, as was done in Ref. \cite{Fraga:2008qn}, one finds that the 
confinement and chiral phase transition lines coincide. Moreover, in this case an increasing magnetic 
field lowers the equivalent chiral-confinement transition temperature. On the other hand, the inclusion 
of the vacuum contribution from quark loops in a magnetic field modifies completely the picture: 
confinement and chiral transition lines split, and both chiral and deconfining 
critical temperatures become increasing functions of the magnetic field. Both scenarios are 
shown in Fig.~\ref{fig:phase:diagram}, which exhibit the full calculation of the phase diagram from 
the effective potential within the \M\ effective model.
The vacuum contribution from the quarks affects drastically the chiral sector as well.
Our calculations also show that the vacuum contribution seems to soften the order of the phase 
transition: the first-order phase transition -- which would be realized in the absence of the vacuum 
contribution -- becomes a smooth crossover in the system with vacuum quark loops included. 

The modifications produced by strong magnetic fields over strong interactions 
seem very exciting, bringing new possibilities for the phase diagram: affecting the nature of 
the transitions, splitting different coexistence lines, possibly exhibiting new phases, 
increasing the breaking of $\Z_3$, and so on. 
As discussed in the Introduction, the second scenario has been also found in other effective 
models containing a chiral and a Polyakov loop sector \cite{Fukushima:2010fe,Skokov:2011ib}, 
as well as in preliminary lattice simulations \cite{lattice-delia}. However, lattice 
simulations of magnetic QCD with physical masses and fine grids have shown that both critical 
temperatures actually go down for increasing $B$, saturating for very large fields \cite{Bali:2011qj}, 
an unexpected behavior that is very different from the scenario depicted above.

This leads us to consider of a much simpler model that, yet, seems to contain the essential 
ingredients to describe the behavior of the deconfining line, and produces results that are in 
qualitative agreement with the lattice: the magnetic MIT bag model \cite{Fraga:2012fs}. 
As will be clear in the 
discussion, the subtraction procedure in renormalization is subtle (which can be seen as the 
choice of the renormalization scale) but can be guided by known physical phenomena and 
lattice results.

\section{Magbag - the thermal MIT bag model in the presence of a magnetic background}

In the MIT bag model framework for the pressure of strong interactions, one needs 
the free quark pressure. As seen previously, the presence of a magnetic field in the ${\bf \hat z}$ 
direction affects this computation by modifying the dispersion relation to 
\begin{equation}
\omega_{\ell sf}(k_{z})=k_{z}^{2}+m_{f}^{2}+q_{f}B(2\ell+s+1)\equiv k_{z}^{2}+M_{\ell sf}^{2} \,,
\label{landau-levels}
\end{equation}
$\ell=0,1,2,\dots$ being the Landau level index, $s=\pm 1$ the spin projection, $f$ the flavor index, 
and $q_{f}$ the absolute value of the electric charge. Loop integrals are also affected as presented 
previously \cite{Fraga:2008qn,Chakrabarty:1996te}.

Since it has been shown that only very large magnetic 
fields do affect significantly the structure of the phase diagram for strong interactions 
\cite{Agasian:2008tb,Fraga:2008qn,Boomsma:2009yk,Fukushima:2010fe,lattice-delia,Bali:2011qj}, 
we can restrict the free quark pressure to the limit of very high magnetic fields, where it is 
possible to simplify some analytic expressions. 

It is crucial to realize, however, that the lowest 
Landau level (LLL) approximation for the free gas pressure is not equivalent to the leading order 
of a large magnetic field expansion. For the zero-temperature, finite-$B$ contribution to the 
pressure, the LLL is the energy level which less contributes in the limit of large $B$; the result 
being dominated by high values of $\ell$. Nevertheless, the equivalence between the LLL 
approximation and the large $B$ limit remains valid for the temperature-dependent part of the 
free pressure (as well as for the propagator), simplifying the numerical evaluation of thermal integrals 
\cite{leticia-tese}. 

The free magnetic contribution to the quark pressure has been considered in different contexts 
(usually, in effective field theories 
\cite{Fraga:2008qn,Menezes:2008qt,Boomsma:2009yk,Andersen:2011ip,Ebert:2003yk}) 
and computed from the direct knowledge of the energy levels of the system, 
Eq. (\ref{landau-levels}). The exact result, including all Landau levels, has to be computed from
\begin{equation}
P_{q}=
2 N_c \sum_{\ell,s,f}\frac{q_f B}{2\pi}
\int \frac{dk_z}{2\pi} \bigg\{ 
\frac{\omega_{\ell sf}(k_z)}{2}
+T \ln\left[ 1+e^{-\omega_{\ell sf}(k_z)/T} \right]
\bigg\}
\, ,\label{bolhaquark}
\end{equation}
where the first term is a clearly divergent zero-point energy and the other one is the 
finite-temperature contribution for vanishing chemical potential. Since $\omega_{\ell sf}$ 
grows with $B$, the largest the $\ell$ labeling the Landau level considered the larger the 
zero-point energy term becomes, being minimal for the LLL, corroborating the previous 
discussion. Thus, in the limit of large $B$, the LLL approximation is inadequate here. 
The decaying exponential dependence of the 
finite-temperature term on $\omega_{\ell sf}$, on the other hand, guarantees that the LLL 
dominates indeed this result for intense magnetic fields. 

To obtain a good approximation for the large $B$ limit of the free pressure, we choose to 
treat the full exact result and take the leading order of a $x_f\equiv m_f^2/2q_fB$ expansion in the 
final renormalized expression. Let us then discuss the treatment of the divergent zero-point 
term. Despite being a zero-temperature contribution, the first term in Eq. (\ref{bolhaquark}) 
cannot be fully subtracted because it carries the modification to the pressure brought about 
by the magnetic dressing of the quarks. Using dimensional regularization and the zeta-function representation, which is also a type of regularization, for the sums over Landau levels and 
subtracting the pure vacuum term in $(3+1)$ dimensions, one arrives at:
\begin{eqnarray}
P_{q}^{V}&=&
\frac{N_c}{2\pi^2} \sum_f (q_fB)^2\Big[
\zeta'\left(-1,x_f\right)
+\frac{1}{2}(x_f-x_f^2)\ln x_f+\frac{x_f^2}{4}
\nonumber \\
&-& \frac{1}{12}\big(
2/\epsilon+\log (\Lambda^2/2q_fB)+1\big)
\Big] \, ,
\end{eqnarray}
where a pole $\sim (q_fB)^2[2/\epsilon]$ still remains. This infinite contribution that survives the 
vacuum subtraction can be interpreted as a pure magnetic pressure coming from the artificial 
scenario adopted, with a constant and uniform $B$ field covering the whole universe (analogous 
to the case of a cosmological constant). In this vein, one may neglect all terms $\sim (q_f B)^2$ 
and independent of masses and other couplings (as done, e.g. in Refs. 
\cite{Fraga:2008qn,Menezes:2008qt,Andersen:2011ip}), concentrating on the modification of the pressure of the quark matter under investigation. This can be seen as a choice for the renormalization scale after 
the renormalization of a $\sim F_{\mu\nu}F^{\mu\nu}$ term representing the magnetic field, 
as discussed, e.g. in Ref. \cite{Andersen:2011ip}. We will come back to this point in the sequel.

The final exact result for the free pressure of magnetically dressed quarks is therefore 
\begin{eqnarray}
\frac{P_{q}}{N_c}&=&
\sum_f \frac{(q_fB)^2}{2\pi^2}\Big[
\zeta'\left(-1,x_f\right)-\zeta'\left(-1,0\right)
+\frac{1}{2}(x_f-x_f^2)\ln x_f+\frac{x_f^2}{4}
\Big] \nonumber \\
&+&T \sum_{\ell,s,f}\frac{q_fB}{2\pi^{2}}
\int dk_{z} 
\ln\left[ 1+e^{-\omega_{\ell sf}(k_z)/T} \right]
\,.
\end{eqnarray}
In Refs. \cite{Fraga:2008qn,Menezes:2008qt,Andersen:2011ip}, the constant 
$\zeta'\left(-1,0\right)=-0.165421...$ was not subtracted. In the case of pions, however, the 
full subtraction ensures that magnetic catalysis, i.e. an enhancement of chiral symmetry breaking, at zero 
temperature \cite{Gusynin:1994xp,Klimenko:1998su,Semenoff:1999xv,Miransky:2002rp,Shovkovy:2012zn}, 
is realized. On the other hand, if this term is left, the pion contribution to the effective potential for the chiral condensate at large magnetic fields will eventually raise the minimum instead of lowering it.

In the limit of large magnetic field (i.e. $x_f=m_f^2/(2q_fB)\to 0$), we obtain
\begin{equation}
\frac{P_{q}}{N_c}\stackrel{{\rm large}~ B}{=}
\sum_f \frac{(q_fB)^2}{2\pi^2}\Big[
x_f\ln \sqrt{x_f}
\Big]
+T \sum_{f}\frac{q_fB}{2\pi^{2}}
\int dk_{z}
\ln\left[ 1+e^{-\sqrt{k_{z}^{2}+m_{f}^{2}}/T} \right]
\,.
\label{quark-pressure-largeB}
\end{equation}

Adding the free piece of the gluonic contribution and the bag constant ${\cal B}$, 
the pressure of the QGP sector in the presence of an intense magnetic field reads:
\begin{equation}
P^{B}_{\rm QGP}=
2(N_c^2-1)\frac{\pi^2T^4}{90}+P_{q}-{\cal B}
\,.
\end{equation}

It is clear that, for $\sqrt{eB}$ much larger than 
all other energy scales, the pressure in the QGP phase increases with 
the magnetic field, which seems to favor a steady drop in the critical temperature with 
increasing $B$ that would lead to a crossing of the critical line with the $T=0$ axis 
at some critical value for the magnetic field. However, the behavior of $T_{c}(B)$ 
also depends on how the pions react to $B$, so that the outcome is not obvious.

In the confined sector, which we describe by a free pion gas, one may follow analogous steps 
in order to compute the contribution from the charged pions, 
which couple to the magnetic field, arriving at
\begin{eqnarray}
P_{\pi^{+}}+P_{\pi^{-}}&=&
-\frac{(e B)^2}{4\pi^2}\Big[
\zeta'\left(-1,\frac{1}{2}+x_{\pi}\right)-\zeta'\left(-1,\frac{1}{2}\right)
+\frac{x_{\pi}^{2}}{4} -x_{\pi}^{2} \ln \sqrt{x_{\pi}}
\Big] \nonumber \\
&-&2 \frac{eB}{4\pi^{2}}
T \sum_{\ell} \int dk_{z} 
\ln\left[ 1-e^{-\sqrt{k_{z}^{2}+M^{2}_{\pi \ell}}/T} \right] \,,
\end{eqnarray}
where $M^{2}_{\pi \ell}\equiv m_{\pi}^{2}+(2\ell +1)eB$ and $x_{\pi}\equiv m_{\pi}^{2}/(2eB)$. In 
this final expression all terms $\sim (q_f B)^2$ and independent of masses and other couplings 
were subtracted, as discussed before. 
Notice that the spin-zero nature of the pions guarantees that {\it all} charged pion modes in a magnetic field, differently from what happens with the 
quark modes, are $B$-dependent. So, in the large magnetic field limit the thermal integral 
associated with $\pi^{+}$ and $\pi^{-}$ is exponentially suppressed by an effective mass 
$\gtrsim (m_{\pi}^2+eB)$, as was also noticed in Ref. \cite{Fraga:2008qn}, and can be dropped. In this 
limit, we have
\begin{equation}
P_{\pi^{+}}+P_{\pi^{-}}\stackrel{{\rm large}~ B}{=}
-\frac{(e B)^2}{4\pi^2} \zeta^{(1,1)}(-1,1/2) ~x_{\pi} \,,
\label{pion-pressure-largeB}
\end{equation}
where $\zeta^{(1,1)}(-1,1/2)=-\ln(2)/2=-0.346574\cdots$. Neutral pions do not couple to the magnetic 
field and contribute only with the usual thermal integral \cite{kapusta-gale}.

As before, for $\sqrt{eB}$ much larger than all other scales, the pion 
pressure rises with the magnetic field, as a consequence of the subtraction of all 
terms that are independent of temperature, masses and other couplings in the 
renormalization process, which renders the pressure positive. Differently from the 
quark pressure, however, the $B=0$ pion pressure takes over for 
temperatures of the order of the pion mass, which is not small and always enlarged 
by the presence of a magnetic field (given its scalar nature). Moreover, for large 
$T$, the magnetic pion pressures converge to $(1/3)$ of the $B=0$ pressure, since 
$\pi^{0}$ is the only degree of freedom that contributes thermally for large $B$.

\begin{figure}[htb]
\begin{center}
\includegraphics[width=7.5cm]{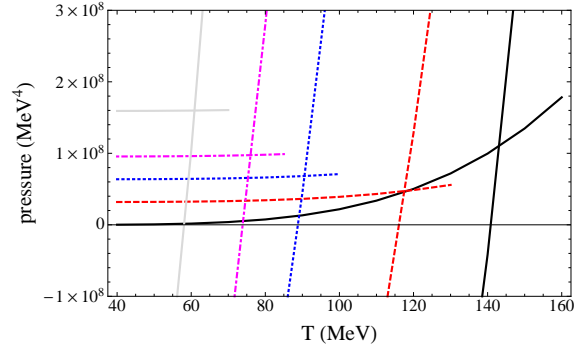}
\end{center}
\caption{Crossing pion gas and QGP pressures as functions 
of the temperature for different values 
of the magnetic field: $eB=0$ (black, solid, right-most), $20m_{\pi}^2,$ $40m_{\pi}^2,$ 
$60m_{\pi}^2$ (magenta, dash-dotted) and $eB=100m_{\pi}^2$ (gray, solid, left-most), 
where $m_{\pi}=138~$MeV is the vacuum pion mass. Extracted from Ref. \cite{Fraga:2012fs}.}
\label{crossingpressures}
\end{figure}

\begin{figure}[htb]
\vspace{1cm}
\begin{center}
\includegraphics[width=7.5cm]{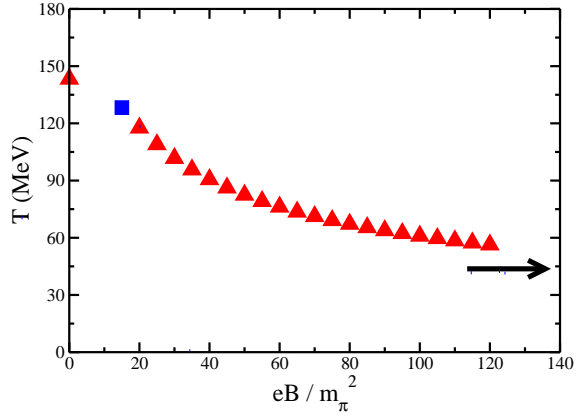}
\end{center}
\caption{Phase diagram in the presence of a strong magnetic field. We also keep the 
$T_{c}(B=0$) point. The blue square represents a very conservative estimate for 
the maximum value of $eB$ expected to be achieved in non-central collisions at the 
LHC with the formation of deconfined matter. 
The arrow marks the critical temperature for $eB\approx 210 m_{\pi}^2$ \cite{Vachaspati:1991nm}, 
expected to be found at the early universe. Extracted from Ref. \cite{Fraga:2012fs}.}
\label{phdiagramB}
\end{figure}

Each equilibrium phase should maximize the pressure, so that the critical line in 
the phase diagram can be constructed by directly extracting $T_{c}(B)$ from the equality 
of pressures. It is instructive, nevertheless, to consider a plot of the crossing pressures, 
as shown in Figure \ref{crossingpressures}. The figure shows, as expected, a decrease 
in the critical temperature (crossing points) as $B$ is increased due to the corresponding 
positive shift of the QGP pressure. However, $T_{c}$ seems to be saturating at a constant value. One can 
see that the critical pressure (crossing point) goes down, 
but then it bends up again due to the increase in the pion pressure with $B$. This 
combination avoids a steady and rapid decrease of the critical temperature, as becomes 
clear in the phase diagram shown in Figure \ref{phdiagramB}. In fact, inspection of the 
zero-temperature limit of 
Eqs. (\ref{quark-pressure-largeB}) and (\ref{pion-pressure-largeB}) 
shows that there is no value of magnetic field that allows for a vanishing 
critical temperature.

The phase diagram in the plane $T-eB$ shows that the critical temperature for deconfinement 
falls as we increase the magnetic field. However, instead of falling with a rate that will 
bring it to zero at a given critical value of $eB$, it falls less and less rapidly, tending to 
saturate at large values of $B$. Remarkably, this qualitative behavior agrees quite well
with the most recent lattice results with physical masses \cite{Bali:2011qj}\footnote{Of course, 
our description necessarily predicts a first-order transition, as usual with 
the MIT bag model, and our numbers should be taken as rough estimates, as is always 
the case in effective models.}. As discussed in the Introduction, previous 
models \cite{Fraga:2008qn,Mizher:2010zb,Fukushima:2010fe,Johnson:2008vna,Preis:2010cq,Skokov:2011ib}, have predicted either an increase or an essentially flat behavior for the deconfinement critical line 
as $B$ is increased to very large values. The same was true for previous lattice 
simulations \cite{lattice-delia}, which could be reproduced by the authors of Ref. \cite{Bali:2011qj} 
by increasing the quark masses to unphysical values. 

The renormalization procedure in the presence of a constant and uniform magnetic field 
seems to be very subtle and crucial for the phenomenological outcome for the phase structure. 
$B$-dependent, mass-independent  terms survive pure vacuum ($B=0$) subtraction and have to 
be subtracted either in an {\it ad hoc} fashion \cite{Fraga:2012fs} or by including a background field 
counterterm associated with a term $\sim F_{\mu\nu}F^{\mu\nu}$ representing the magnetic 
field \cite{Andersen:2011ip}. The latter brings a renormalization scale and, upon an appropriate 
choice, reproduces the former. Subtracting all purely magnetic terms in the pressures seems to be 
the appropriate choice since: (i) one guarantees that the pion pressure grows with increasing magnetic 
field at zero temperature, which is consistent with the well-known phenomenon of magnetic catalysis; 
(ii) lattice simulations usually measure derivatives of the pressure with respect to temperature and quark 
mass, and do not access derivatives with respect to $B$, so that purely $B$-dependent terms are not 
included in their results; and (iii) the effect of a purely magnetic contribution to the pressure 
would only shift the effective potential as a whole. In particular, there would be no modification on 
relative positions and heights of different minima that represent different phases of matter.

The qualitative success of the description of the deconfinement transition in the presence of an 
external magnetic field in terms of the MIT bag model suggests that confinement dynamics plays 
a central role in guiding the functional behavior of $T_{c}$. In this case, a large $N_{c}$ investigation 
of the associated magnetic thermodynamics seems appropriate.

\section{Large $N_{c}$}

Lattice QCD calculations \cite{Teper:2008yi} show that the deconfinement phase transition of pure glue $SU(N_c)$ gauge theory becomes first order when 
$N_c \geq 3$ \cite{Boyd:1996bx,muitaslattice,Lucini:2005vg,Borsanyi:2012ve} with a critical 
temperature given by \cite{Lucini:2012wq}
\begin{equation}
\lim_{N_c\to\infty}\frac{T_c}{\sqrt{\sigma}}=0.5949(17) + \frac{0.458(18)}{N_c^2} \, ,
\end{equation}
where $\sigma \sim (440\, {\rm MeV})^2$ is the string tension. 
The thermodynamic properties of pure glue do not seem to change appreciably when $N_c \geq 3$ \cite{morelattice,Panero:2009tv}, which suggests that large $N_c$ arguments may indeed capture the 
main physical mechanism behind the deconfinement phase transition of QCD. 

It has been shown in Ref. \cite{mag-largeNc} that the deconfinement critical temperature must decrease 
in the presence of an external magnetic field in the large $N_c$ limit of QCD, provided that quarks exhibit 
a paramagnetic behavior. Assuming that $N_f/N_c \ll 1$ and $m_q=0$, the only contribution 
to the pressure of the confined phase that enters at $\mathcal{O}(N_c^2)$ is given by the vacuum 
($B=0$) gluon condensate $c_0^4 N_c^2 \sigma^2$. 
The gluon and quark condensates change in the presence of a magnetic field \cite{Agasian:1999sx,Cohen:2007bt,Agasian:2000hw} but these modifications are negligible in the large $N_c$ limit. 
Besides, the gluon contribution to the deconfined pressure is blind to the magnetic field. 

On the other hand, the quark contribution is affected by the magnetic field and has the form 
\begin{equation}
P_{quark}(T,eB)\sim N_c \,N_{pairs}(N_f) \,T^4\,\tilde{f}_{quark}(T/\sqrt{\sigma},eB/T^2)\,,
\end{equation}
with $N_{pairs}(N_f)/N_c \ll 1$ being the number of pairs of quark flavors with electric charges $\left\{ (N_{c}-1)/N_{c},-1/N_{c}\right\}$ in units of the fundamental charge. Only the largest ($\sim N_{c}^{0}$) charge in each pair contributes to leading order in $N_{f}/N_{c}$.
Notice that the function $\tilde{f}_{quark}$ is positive definite and must increase monotonically with $T$ for a fixed value of $eB$ until it goes to 1 in the high temperature limit $T \gg \sqrt{\sigma}$, $eB$. Thus, one should expect that the critical temperature as a function of the magnetic field, $T_c(eB)$, must decrease with respect to the pure glue critical temperature, $T_c^{(0)}$, by an amount of $\mathcal{O}(N_f/N_c)$. This can be seen directly by equating the pressures 
at $T_c$, which yields \cite{mag-largeNc}
\begin{equation}
\frac{T_c(eB)}{\sqrt{\sigma}}\,f_{glue}^{1/4}\left(\frac{T_c(eB)}{\sqrt{\sigma}}\right)=\frac{c_2(N_{pairs},eB)}{c_{SB}}\,,
\label{newTcmag}
\end{equation}
where we defined
\begin{eqnarray}
c_2(N_{pairs},eB) \equiv c_0 \left[1-\frac{1}{4}\frac{N_{pairs}(N_f)}{N_c}\frac{c_{q\,SB}^4\,\tilde{f}_{quark}\left(\frac{T_c^{(0)}}{\sqrt{\sigma}},\frac{eB}{T_c^{(0)\,2}}\right)}
{c_{SB}^4\,f_{glue}\left(\frac{T_c^{(0)}}{\sqrt{\sigma}}\right)}\right]\,.
\label{c2B}
\end{eqnarray}   
\begin{figure}[h!]
\begin{center}
\includegraphics[width=6.0cm,angle=90]{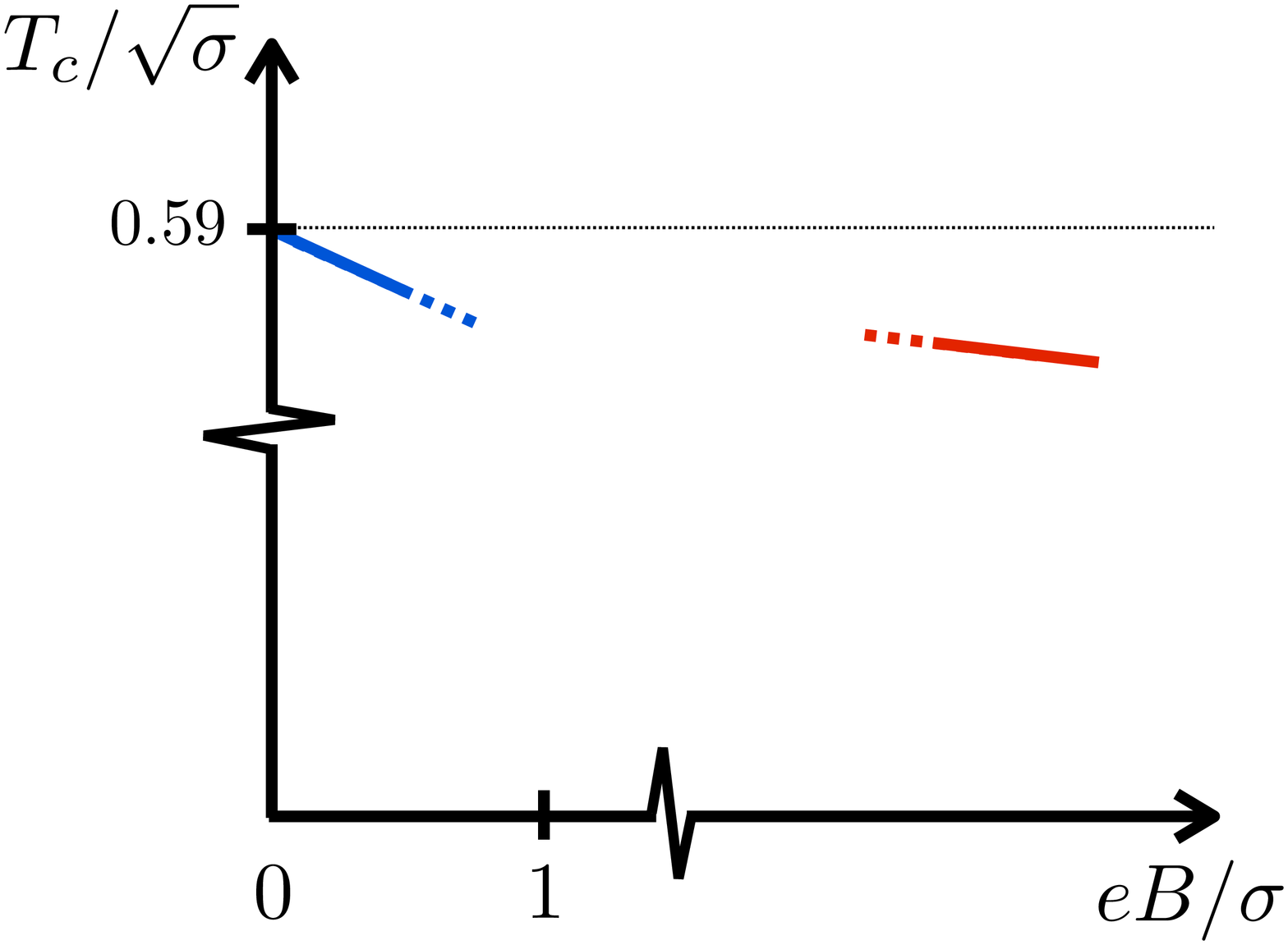}
\end{center}
\caption{Cartoon of the $T_c \times eB$ phase diagram in the large $N_{c}$ limit, using the approximation of 
free deconfined quarks and gluons. The numerical value $0.59$ shown in the plot was extracted from Ref. \cite{Lucini:2012wq}. Extracted from Ref. \cite{mag-largeNc}.}
\label{cartoon-largeN}
\end{figure}

Since $c_2(N_{pairs},eB)< c_0$, one finds that $T_c(eB)/T_c^{(0)}< 1$ by an amount 
$\sim N_f/N_c$ \cite{mag-largeNc}. Assuming that quarks behave paramagnetically for all values of $B$, 
then $c_2(N_{pairs},eB)< c_1(N_{f})$, its equivalent in the case with $B=0$ and $N_{f}>0$, and $T_c(eB)$ is 
also lower than the critical temperature in the presence of $N_{f}/N_{c}$ flavors of  massless quarks 
at $B=0$ \cite{mag-largeNc}. 

In a free gas implementation of the deconfined phase $f_{glue}=1$ and, for very strong magnetic fields, 
$\tilde{f}_{quark}\sim eB/T_c^2$ \cite{Fraga:2012fs}, so that the magnetic suppression of the deconfinement critical temperature goes like $eB\,N_{pairs}/(N_c\,\sigma)$. This simple implementation in the limits of low and high magnetic fields provides a scenario in which the 
slope in $T_c(eB)$ decreases for large fields, as illustrated in Fig. \ref{cartoon-largeN}, which hints 
for a saturation of $T_c$ as a function of $eB$, as observed on the lattice \cite{Bali:2011qj} and in the 
magnetic MIT bag model \cite{Fraga:2012fs}, but cannot be obtained in a model-independent way.

\section{Conclusions and perspectives}

The investigation of the effects brought about by the presence of a magnetic background on the 
thermal chiral and deconfining transitions is in its infancy yet. Nevertheless, the promise of the outcome 
of a rich phenomenology in mapping this new phase diagram of strong interactions is concrete.

First model calculations have revealed the possibility of modifications in the nature of the QCD phase 
transitions, and also the appearance of a new phase of strong interactions in the case of a splitting of the 
critical (chiral and deconfining) lines. Even if recent, more physical lattice simulations have drastically modified the initial picture, 
they have also shown that the magnetic background has a very non-trivial influence on strong interactions. 
For instance, the behavior of quark condensates at finite temperature is non-monotonic \cite{Bali:2012zg}, 
rendering well-established vacuum phenomena such as magnetic catalysis more subtle at finite temperature.

The functional behavior of the critical temperatures still has to be understood more deeply. Although no 
model foresaw the fact that both, chiral and deconfining temperatures, decrease then saturate at a nonzero 
value according to the lattice \cite{Bali:2011qj}, {\it a posteriori} the magnetic MIT bag model was successful 
to describe this behavior for deconfinement qualitatively \cite{Fraga:2012fs} and seems to capture some 
essential ingredients. A model-independent analysis in the large-$N_{c}$ limit of QCD also points to this 
behavior \cite{mag-largeNc}, which is reassuring from the theoretical standpoint.

Another key ingredient in building an understanding of the physics of the quark-gluon plasma under these 
new conditions, which can be relevant for high-energy heavy-ion collision experiments, the primordial 
quark-hadron transition and magnetars is the standard perturbative investigation of magnetic QCD. 
The calculation of the pressure in thermal QCD to two loops 
in the strong sector using the full QED propagator in the lowest Landau level approximation is subtle but 
possible, as done originally in Ref. \cite{leticia-tese}. 

The computation makes use of the full 
magnetic propagator that was obtained by Schwinger \cite{Schwinger:1951nm}, but can 
be cast in a more convenient form in terms of a sum over Landau levels as derived in 
Ref. \cite{Chodos:1990vv} (see also Refs. \cite{leticia-tese,Fukushima:2011nu}). In particular, it has been 
shown in Ref. \cite{leticia-tese} that the chiral limit for the exchange diagram seems to be trivial for very 
large magnetic fields. Concretely, it can be written diagrammatically in the following compact 
form \cite{leticia-tese}:

\begin{fmffile}{fmftese}

\begin{eqnarray}
\parbox{10mm}{
\begin{fmfgraph*}(35,35)\fmfkeep{exchange}
\fmfpen{0.8thick}
\fmfleft{i} \fmfright{o}
\fmf{fermion,left,tension=.08}{i,o,i}
\fmf{gluon}{i,o}
\fmfdot{i,o}
\end{fmfgraph*}
}^{\quad\rm LLL}
 &=& 
\left(
\frac{q_fB}{2\pi}\right)
\int\frac{dk_1dk_2}{(2\pi)^2}~{\rm e}^{
-\frac{k_1^2+k_2^2}{2q_fB}}\quad
\parbox{10mm}{
\fmfreuse{exchange}
}^{\quad  \bar d=2;\; m_k^2=k_1^2+k_2^2}
\, ,
 \label{excRF-LLL-4}
\end{eqnarray}

\end{fmffile}

\noindent
which realizes the intuitive expectation that the nontrivial dynamics in an extremely intense magnetic field should be one-dimensional. Since gluons do not couple directly to the magnetic field, their dispersion relation maintains its three-dimensional character, which effectively results in a ``massive'' gluon in the dimensionally-reduced diagram. In the end the exchange contribution to the pressure is essentially an average over the effective gluon transverse mass $m_k^2=k_1^2+k_2^2$ of the exchange diagram in $(1+1)$-dimensions with the Gaussian weight $(q_fB/2\pi) \exp[-m_k^2/2q_fB]$. Since the trace in the reduced diagram is proportional to $m_{f}^{2}$, the chiral limit seems trivial \cite{leticia-tese}.
A detailed analysis of the dependence of the pressure on the mass and temperature and a semiclassical 
interpretation of this result will be reported soon \cite{magnetic-pressure}. 

The nature of the phase diagram of strong interactions in the presence of a magnetic background is 
still open. Recent lattice data, especially when compared to effective model predictions, seem to indicate 
that confinement dynamics plays an important role in the phase structure that emerges and should be 
incorporated in any effective description. Comparison between lattice data with very different quark masses 
\cite{lattice-delia,Bali:2011qj,Bali:2012zg} also show that the dependence of the critical temperatures on 
this parameter is non-trivial: $T_{c}$ increases at the percent level for large masses \cite{lattice-delia} 
whereas it decreases appreciably for physical masses \cite{Bali:2011qj}. This competition between 
the effects from the magnetic field and quark masses on $T_{c}$ was also found in the large-$N_{c}$ 
QCD analysis of Ref. \cite{mag-largeNc}. A more systematic analysis of this phenomenon on the 
lattice would be very helpful for the building of effective models.

\begin{acknowledgement}
It is a pleasure to thank the editors of this volume of {\it Lecture Notes in Physics} for the 
invitation to contribute to this special edition. The discussion presented here in part summarizes 
work done in collaboration with M.N. Chernodub, A.J. Mizher, J. Noronha and L.F. Palhares to 
whom I am deeply grateful. I am especially indebted to my former students 
A.J. Mizher and L.F. Palhares, from whom I have learnt so much over several years. 
I also thank J.-P. Blaizot, M. D'Elia and G. Endrodi for fruitful discussions.
\end{acknowledgement}

\end{document}